\documentclass[reqno,a4paper]{revtex4}

\makeindex

\usepackage{amsmath,graphicx,mathrsfs}

\newtheorem{prop}{Proposition}

\usepackage{amssymb}
\usepackage{eucal}

\allowdisplaybreaks

\def\openone{\leavevmode\hbox{\small1\kern-3.3pt\normalsize1}}

\def\openone{\leavevmode\hbox{\small1\kern-3.3pt\normalsize1}}

\def\openone{\leavevmode\hbox{\small1\kern-3.3pt\normalsize1}}

\def\on#1#2{\mathop{\vbox{\ialign{##\crcr\noalign{\kern2pt}
$\scriptstyle{#2}$\crcr\noalign{\kern2pt\nointerlineskip}
\kern-2pt$\hfil\displaystyle{#1}\hfil$\crcr}}}\limits}

\begin{document}

\arraycolsep=2pt
\bibliographystyle{apsrev}
\title{New Lax pair for restricted multiple three wave interaction
system, quasiperiodic solutions and bi-hamiltonian structure}

\author{N.\ A.\ Kostov$^1$, A.\ V.\ Tsiganov$^{2}$}

\affiliation{$^1$Institute for Electronics, Bulgarian Academy of
Sciences, \\ Blvd. Tzarigradsko chaussee 72, 1784 Sofia, Bulgaria}

\affiliation{$^2$ Department of Mathematical and Computational Physics\\
St Petersburg State University, Russia}

\date{\today}

\begin{abstract}
We study restricted multiple three wave interaction system by the
inverse scattering method. We develop the algebraic approach in
terms of classical $r$-matrix and give an interpretation of the
Poisson brackets as linear $r$-matrix algebra. The solutions are
expressed in terms of polynomials of theta functions. In particular
case for $n=1$ in terms of Weierstrass functions.
\end{abstract}
 \maketitle

\section{Restricted multiple three wave interaction system}
Several studies have appeared recently on coupled quadratic
nonlinear oscillators \cite{mcl82},\cite{mcl83},\cite{wjb86}
\begin{eqnarray}
&&\imath \frac{d b_{j}}{d\xi}+u c_{j} -\frac{1}{2} \epsilon_{j}
b_{j}=0,
\label{mod1}\\
&&\imath \frac{d c_{j}}{d\xi}+u^{*} b_{j} +\frac{1}{2} \epsilon_{j}
c_{j}=0,
\label{mod2} \\
&&\imath \frac{d u}{d\xi}+\sum_{j=1}^{n}b_{j}c_{j}^{*}=0,
\label{mod3}
\end{eqnarray}
where $\xi$ is the evolution coordinate and $\epsilon_{j}$
corresponds to the normalized wave number mismatchs. The system
(\ref{mod1}-\ref{mod3}) is introduced in \cite{wwc76} to model the
growth of a low frequency internal ocean wave by interaction with
higher frequency surface waves and is used in \cite{m79} as a model
of plasma turbulence. This system describe triads of waves $(a_j,
b_j, u)$, $j=1,\ldots n$ evolving in $\xi$ and interacting with each
other trough multiple three wave interaction with possible
applications in optics.

\section{Lax representation}
Let us consider coupled quadratic nonlinear oscillators
\begin{eqnarray}
&&\imath \frac{d b_{j}}{d\xi}+u c_{j} -\frac{1}{2} \epsilon_{j}
b_{j}=0,
\label{mod12}\\
&&\imath \frac{d c_{j}}{d\xi}+u^{*} b_{j} +\frac{1}{2} \epsilon_{j}
c_{j}=0,
\label{mod22} \\
&&\imath \frac{d u}{d\xi}+\sum_{j=1}^{n}b_{j}c_{j}^{*}=0,
\label{mod32}
\end{eqnarray}
where $\xi$ is the evolution coordinate and $\epsilon_{j}$ are
constants. The equations (\ref{mod12}-\ref{mod32})  can be written
as Lax representation
\begin{equation}\label{eq:LAX}
\frac{d L}{d\xi}= [M,L],
\end{equation}
of the following linear system:
\begin{eqnarray}
\frac{d \psi}{d\xi} = M(\xi,\lambda)\psi(\xi,\lambda) \quad
L(\xi,\lambda)\psi(\xi,\lambda)= 0 , \label{lineq}
\end{eqnarray}
where $L,M$ are $2\times2$ matrices and have the form
\begin{eqnarray}\label{Lax}
&&L(\xi,\lambda) = \left( \begin{array}{cc} A(\xi,\lambda)
 &  B(\xi,\lambda)
\\ C(\xi,\lambda) &
D(\xi,\lambda) \end{array} \right) ,
\\&&M(\xi,\lambda)  = \left( \begin{array}{cc} -\imath\lambda/2 & i u
\\ u^{*}
& \imath\lambda/2 \end{array} \right) .
\end{eqnarray}
where
\begin{eqnarray}
&&A(\xi,\lambda) =a(\lambda)\left( -\imath \frac{\lambda}{2}
+\frac{\imath}{2}\sum_{j=1}^{n}
\frac{\left(c_{j}c_{j}^{*}-b_{j}b_{j}^{*}\right) }
{\lambda-\epsilon_{j}}\right), \\ &&B(\xi,\lambda) =a(\lambda)\left(
\imath u -\imath\sum_{j=1}^{n} \frac{b_{j}c_{j}^{*} }
{\lambda-\epsilon_{j}}\right), \\
&&C(\xi,\lambda) =  a(\lambda)\left(\imath u^{*}
-\imath\sum_{j=1}^{n} \frac{c_{j}b_{j}^{*} }
{\lambda-\epsilon_{j}}\right),
\end{eqnarray}
where $D(\xi,\lambda)=-A(\xi,\lambda)$ and
$a(\lambda)=\prod_{i=1}^{n} (\lambda-\epsilon_{i})$. The Lax
representation yields the hyperelliptic curve $K=(\nu,\lambda)$
\begin{equation}
\mbox{det}(L(\lambda)-\frac{1}{2}\nu{\mathrm I})=0 , \label{Lcurv}
\end{equation}
where ${\mathrm I}$ is the $2\times 2$ unit matrix. The moduli of
the curve (\ref{Lcurv}) generate the integrals of motion
$J_{0},J_{j},K_{j}, j=1,\ldots,n$, \begin{equation}
\nu^{2}=A^{2}(\xi,\lambda)+B(\xi,\lambda) C(\xi,\lambda).
\label{Lcurv1}
\end{equation}
The curve (\ref{Lcurv1}) can be written in canonical form as
\begin{eqnarray}
\nu^2=4\prod_{j=1}^{2n+2}(\lambda-\lambda_{j})=R(\lambda),\label{curva1}
\end{eqnarray}
where $\lambda_{j}\neq \lambda_{k}$ are branching points. From
(\ref{Lcurv1}) and explicit expressions for $A(\xi,\lambda),
B(\xi,\lambda), C(\xi,\lambda)$ we obtain
\begin{equation}
\nu^{2}=a(\lambda)^{2}\left(\lambda^2 -4 i J_{0}+4 i \sum_{j=1}^{n}
\frac{K_{j}}{\lambda-\epsilon_{j}} - i
\sum_{j=1}^{n}\frac{J_{j}^{2}}{(\lambda-\epsilon_{j})^{2}}\right),
\label{curven1}
\end{equation}
where
\begin{eqnarray}
&&K_{j}=i ub_{j}^{*} c_{j} + i u^{*}b_{j}
c_{j}^{*}+i\frac{\epsilon_{j}}{2}(|c_{j}|^2-|b_{j}|^2)   \nonumber \\
&&-i\frac{1}{2}\sum_{k\neq j}\frac{
\left((|b_{j}|^2-|c_{j}|^2)(|b_{k}|^2-|c_{k}|^2)+
2b_{j}^{*}c_{j}b_{k}c_{k}^{*}+ 2b_{j}c_{j}^{*}b_{k}^{*}c_{k}
\right) }{\epsilon_{j}-\epsilon_{k}}   ,   \label{int-mot} \\
&&J_{0}=i |u|^2+\frac{1}{2}i\sum_{j}^{n} (|b_{j}|^2-|c_{j}|^2),
\quad J_{j}=i(|b_{j}|^{2}+|c_{j}|^2). \nonumber
\end{eqnarray}
Next we develop a method which allows to construct quasi-periodic
and periodic solutions of system (\ref{mod12}-\ref{mod32}). The
method is based on the application of spectral theory for
self-adjoint one dimensional Dirac equation with quasi-(periodic)
finite gap potential ${\mathcal U}=-u$ cf. Eqs.
(\ref{mod12},\ref{mod22})
\begin{eqnarray}
&&\imath \frac{d \Psi_{1j}}{d\xi}-{\mathcal U} \Psi_{2j}
-i\lambda_{j}\Psi_{1j}=0,
\label{dirac1}\\
&&\imath \frac{d \Psi_{2j}}{d\xi}-{\mathcal U^{*}} \Psi_{1j}
+i\lambda_{j}\Psi_{1j}=0, \label{dirac2}
\end{eqnarray}
with spectral parameter $\lambda$ and eigenvalues
$\lambda_{j}=i\epsilon_{j}/2$. The equation (\ref{eq:LAX}) is
equivalently written as
\begin{eqnarray}
&&\frac{d A}{d \xi} = i u C- i u^{*} B, \quad
A(\xi,\lambda)=\sum_{j=0}^{n+1} A_{n+1-j}(\xi)\lambda^{j},
\label{eq:A}\\
&&\frac{d B}{d \xi} = -i \lambda B - 2 i u A, \quad
B(\xi,\lambda)=\sum_{j=0}^{n} B_{n-j}(\xi)\lambda^{j},
\label{eq:B}\\
&&\frac{d C}{d \xi} = i \lambda C + 2 i u^{*} A, \quad
C(\xi,\lambda)=\sum_{j=0}^{n} C_{n-j}(\xi)\lambda^{j},  \label{eq:C}
\end{eqnarray}
or in different form we have
\begin{eqnarray}
&&A_{j+1,\xi} = i u C_{j} - i u^{*} B_{j}, \, A_{0}=1, A_{1}=c_{1}, \\
&&i B_{j+1} = - B_{j,\xi} - 2 i u A_{j+1},  \quad B_{0}=-2 u,  \\
&&i C_{j+1} =  C_{j,\xi} - 2 i u^{*} A_{j+1} \quad C_{0}=-2 u^{*},
\end{eqnarray}
where $c_{1}$ is the constant of integration. Differenciating Eq.
(\ref{eq:A}) and using (\ref{Lcurv1}) we can obtain
\begin{eqnarray}
B B_{\xi\xi} -\frac{u_{\xi}}{u} B B_{\xi} -\frac{1}{2} B_{\xi}^{2}+
\left(\frac{\lambda^{2}}{2}-i\lambda\frac{u_{\xi}}{u} +
|u|^{2}\right) B^{2} =2 u^{2} \nu.
\end{eqnarray}

Using (\ref{Psi1}) the eigenfunction $\Psi_{1}$ for finite-gap
potential ${\mathcal U}$ have the form

\begin{eqnarray}
\Psi_{1}(\xi,\lambda)=\left[\frac{{\mathcal U(\xi)}}{{\mathcal
U(0)}}
\prod_{j=1}^{n}\frac{\lambda-\mu_{j}(\xi)}{\lambda-\mu_{j}(0)}\right]^{1/2}
\exp\left\{-i\int_{0}^{\xi}\frac{\sqrt{R(\lambda)}}{\prod_{j=1}^{n}(\lambda-\mu
(\xi'))}d\xi'\right\}. \nonumber
\end{eqnarray}

Special case of system (\ref{mod12}-\ref{mod32}) is the three wave
system
\begin{eqnarray}
\imath \frac{d A_{1}}{d\xi}= \epsilon A_{3} A_{2}^{*} , \quad \imath
\frac{d A_{2}}{d\xi}= \epsilon A_{1}^{*} A_{3}, \quad \imath \frac{d
A_{3}}{d\xi}= \epsilon A_{1} A_{2}. \label{3mod}
\end{eqnarray}
The corresponding elements of Lax matrices are
\begin{eqnarray}\label{3wave}
&&L(\xi,\lambda) = \left( \begin{array}{cc} -\imath\frac{\lambda}{2}
+\frac{\imath}{2\lambda} {\mathcal A}&  -i\epsilon A_{1} -
i\frac{1}{\lambda} A_{3}A_{2}^{*} \\ -i\epsilon
A_{1}^{*}-i\frac{1}{\lambda} A_{2}A_{3}^{*2} &
\imath\frac{\lambda}{2} -\frac{\imath}{2\lambda} {\mathcal A}
\end{array} \right) ,
\\&&M(\xi,\lambda)  = \left( \begin{array}{cc}
-\imath\frac{\lambda}{2} & -i\epsilon A_{1} \\ -i\epsilon A_{1}^{*}
& -\imath{\lambda}{2} \end{array} \right),\quad {\mathcal
A}=|A_{2}|^2-|A_{3}|^2 .\end{eqnarray}

To integrate the system  (\ref{3mod}) we introduce new variable
\begin{equation} \label{3mudef}
\mu=\imath \frac{1}{A_{1}} \frac{dA_{1}}{d\xi} =
\frac{A_{3}A_{2}^{*}}{A_{1}},
\end{equation}
in terms of which our equations can be written as
\begin{equation}
\frac{d\mu}{d\xi} = 2\imath\sqrt{R(\mu)},
\end{equation}
where
\begin{eqnarray}&&R(\lambda)=(\frac{1}{4}\lambda^{2}
-\frac{1}{2}{\mathcal A})^{2}
-|A_{1}|^{2}(\lambda-\mu)(\lambda-\mu^{*})= \\ &&
\frac{1}{4}\lambda^{4}-\alpha_{1}\lambda^{3}+\alpha_{2}\lambda^{2}-
\alpha_{3}\lambda+\alpha_{4}.\end{eqnarray} The $\mu$ variable and
$A_{j}, j=1,\ldots 3$ obey the equations
\begin{eqnarray}
&&\alpha_{1}=0, \quad  |A_{1}|^{2} - \frac{1}{2}{\mathcal A} =
\alpha_{2}, \\ &&  - |A_{1}|^{2}(\mu + \mu^{*}) = \alpha_{3}, \quad
\frac{1}{4}{\mathcal A}-\mu\mu^{*} |A_{1}|^{2} = \alpha_{4}.
\label{3eqsmu}\end{eqnarray} which are related to the integrals of
motion of the monomer system with $\alpha_{4}=0$.The equation of
motion is then \begin{eqnarray}
\left(\frac{d\mu}{d\xi}\right)^{2}=-4\left(\mu^{4}
+N\mu^2-H\mu\right)\end{eqnarray} where the system (\ref{3mod})
conserves the dimensionless variable $N$ and the Hamiltonian
$H$\begin{equation} N=|A_{1}|^{2}-\frac{1}{2}{\mathcal A},\quad
H=A_{1}A_{2}A_{3}^{*}+A_{3}A_{1}^{*}A_{2},
\end{equation}
Solving Eqs. (\ref{3eqsmu}) for $\mu$ variable we obtain
\begin{eqnarray} \label{3muvar}
\mu=\frac{1}{4\nu}\left(H+\imath\sqrt{P(\nu)}\right),\end{eqnarray}
where
\begin{equation}
P(\nu)=4\nu^3-4N\nu^2+ N^2\nu-H^2 .
\end{equation}
We seek the solution $A_{1}$ in the following form
\begin{eqnarray}
A_{1}=\sqrt{\nu(\xi)}\exp\left(
 i C\int_{0}^{\xi}\frac{d\xi'}{\nu(\xi')}\right)=
\sqrt{\nu(\xi)}\exp(i\psi(\xi)) , \label{3vv}\end{eqnarray} where
$\nu=|A_1|^2=\wp(\xi+\omega')+C_1$, $\wp$ is the Weierstrass
function, and $\omega'$ is half period.Using Eq. (\ref{3muvar}) and
the following equation
\begin{eqnarray}
\frac{d\nu}{d\xi}=-2\imath\nu(\mu-\mu^{*}),\end{eqnarray} derived
from (\ref{3mudef}) and three wave equations we obtain
\begin{eqnarray}
\left(\frac{d\nu}{d\xi}\right)^2=4\nu^3-4N\nu^2+ N^2\nu-H^2,
\end{eqnarray} whose solution can be expressed in terms of the
Weierstrass elliptic functions as
\begin{eqnarray}
\nu=\wp(\xi+\omega')+\frac{N}{3}.
\end{eqnarray}
Substituting this expression in Eq. (\ref{3vv}) we obtain
\begin{eqnarray}
A_1=\sqrt{\wp(\xi+\omega')+\frac{N}{3}}\,\exp(i\psi(\xi)),
\end{eqnarray}
where the phase $\psi(\xi)$ is given by
\begin{eqnarray}
\psi(\xi)=\frac{H}{2\wp'(\kappa)}\left(\mbox{ln}
\frac{\sigma(\xi+\omega'-\kappa)}{\sigma(\xi+\omega'+\kappa)}+
2\zeta(\kappa)\xi\right) + \psi_0,
\end{eqnarray}
and $\psi_0$ is initial constant phase.

\section{Bi-hamiltonian structure}
In this paragraph we will compute $r$-matrix algebra of restricted
multiple three wave interaction system. We note that in Lax
representation (\ref{Lax}) we remove the function $a(\lambda)$,
which is essential for studying Hamiltonian dynamics of restricted
multiple three wave interaction system.

Let as consider Lax matrix
\begin{equation}\label{lax-a}
{ \mathscr L}(\lambda)=a^{-1}(\lambda) L(\xi,\lambda)=
\left(
\begin{array}{cc}
\mathscr A & \mathscr B \\
\mathscr C & -\mathscr A
\end{array}
\right)\,.
\end{equation}
and introduce standard Poisson
bracket, $\{\cdot,\cdot\}_0$
\begin{eqnarray*}
&&\{f,g\}_0= -\imath\left(\frac {\partial f}{\partial u}
\frac{\partial g}{\partial u^*} -\frac {\partial f}{\partial u^*}
\frac{\partial g}{\partial u}\right)
-\imath\sum_{j=1}^{n}\left(\frac {\partial f}{\partial b_j}
\frac{\partial g}{\partial b_j^*} -\frac {\partial f}{\partial
b_j^*}\frac{\partial g}{\partial b_j}\right) \nonumber \\
&&-\imath\left(\frac {\partial f}{\partial c_j}\frac{\partial
g}{\partial c_j^*} -\frac {\partial f}{\partial c_j^*}\frac{\partial
g}{\partial c_j}\right) \end{eqnarray*}

The entries of $\mathscr L$ satisfy to the following well known equations
\begin{eqnarray*}
\{\mathscr A(\lambda),\mathscr A(\mu)\}_0&=&\{\mathscr B(\lambda),\mathscr B(\mu)\}_0=\{\mathscr C(\lambda),\mathscr C(\mu)\}_0=0,
\cr
\cr
\{\mathscr A(\lambda),\mathscr B(\mu)\}_0&=&\frac{1}{\lambda-\mu}
\Bigl(\mathscr B(\mu)-\mathscr B(\lambda)\Bigr),\cr
\cr
\{\mathscr A(\lambda),\mathscr C(\mu)\}_0&=&\frac{-1}{\lambda-\mu}
\Bigl(\mathscr C(\mu)-\mathscr C(\lambda)\Bigr),\cr
\cr
\{\mathscr B(\lambda),\mathscr C(\mu)\}_0&=&\frac{2}{\lambda-\mu}\Bigl(\mathscr A(\mu)-\mathscr A(\lambda)\Bigr),
\end{eqnarray*}
which may be rewritten as linear $r$-matrix algebra
\begin{equation}
\{\,\on{\mathscr L}{1}(\lambda),\,\on{\mathscr L}{2}(\mu)\}_0= [r(\lambda-\mu),\,
\on{\mathscr L}{1}(\lambda)+\on{\mathscr L}{2}(\mu)\,]\,, \label{rrpoi}
\end{equation}
here $\on{\mathscr L}{1}(\lambda)=\mathscr L(\lambda)\otimes \mathrm I\,,~\on{\mathscr L}{2}(\mu)=\mathrm I\otimes \mathscr L(\mu)$ and
$r(\lambda-\mu)$ is a classical rational $r$-matrix:
\[
r(\lambda-\mu)=\frac{\Pi}{\lambda-\mu},\qquad
\Pi=\left(\begin{array}{cccc}1&0&0&0\\ 0&0&1&0\\ 0&1&0&0\\ 0&0&0&1
\end{array}\right). \]

Remind, that  two  Poisson brackets $\{.,.\}_{0}$ and $\{.,.\}_{1}$ are compatible if every linear combination of them is still a Poisson bracket. The corresponding compatible Poisson tensors $P_0$ and $P_1$ satisfy to the following equations
\begin{equation}\label{sch-eq}
[\![P_0,P_0]\!]=[\![P_0,P_1]\!]=[\![P_1,P_1]\!]=0,
\end{equation}
where $[\![.,.]\!]$ is the Schouten bracket \cite{mag97}. Remind that
on a smooth finite-dimensional manifold $\mathscr M$
the Schouten bracket of two bivectors  $X$ and $Y$  is an antisymmetric contravariant tensor of rank three and its components in local coordinates  $z_m$ read
\[
[\![X,Y]\!]^{ijk}=-\sum_{m=1}^{dim\, \mathscr M} \left(
X^{mk}\dfrac{\partial Y^{ij}}{\partial z_m}+Y^{mk}\dfrac{\partial
X^{ij}}{\partial z_m}+\mbox{cycle}(i,j,k)\, \right).
\]
The Poisson bracket associated with the Poisson bivector $P$ is equal to
\begin{equation} \label{pois-br}
 \{f(z),g(z)\}=\langle df,\,P\,dg \rangle=\sum_{i,k}
P^{ik}(z)\dfrac{\partial f(z)}{\partial z_i}\dfrac{\partial
g(z)}{\partial z_k}\,.
\end{equation}
Here $df$ is covector with entries $\partial f/\partial z_i$ and
$\langle .,. \rangle$ is a standard vector product.

There are a lot of the Poisson brackets
$\{.,.\}_1$ compatible with the linear $r$-matrix bracket (\ref{rrpoi}) similar to the quadratic Sklyanin algebra \cite{ts07a}. Here we consider two examples only.

\begin{prop}
If
\begin{equation}\label{abc1}
\mathscr A=\sum_{i=1}^n  \frac{h_i}{\lambda-\epsilon_i},\quad
\mathscr B=1+\sum_{i=1}^n \frac{e_i}{\lambda-\epsilon_i},\quad
\mathscr C=\sum_{i=1}^n \frac{f_i}{\lambda-\epsilon_i},
\end{equation}
where $h_i,e_i,f_i$ are dynamical variables and $\epsilon_i$ are numerical parameters, then
the following brackets are compatible with linear $r$-matrix bracket (\ref{rrpoi})
\begin{eqnarray}
\{\mathscr B(\lambda),\mathscr B(\mu)\}_1&=&\{\mathscr A(\lambda),\mathscr A(\mu)\}_1=0,\nonumber \\
\nonumber \\
\{\mathscr A(\lambda),\mathscr B(\mu)\}_1&=&\frac{1}{\lambda-\mu}\Bigl(\lambda \mathscr B(\mu)-\mu \mathscr B(\lambda)\Bigr)-\mathscr B(\lambda)\mathscr B(\mu),\nonumber \\
\nonumber \\
\{\mathscr A(\lambda),\mathscr C(\mu)\}_1&=&\frac{-\lambda}{\lambda-\mu}\Bigl(\mathscr C(\mu)-\mathscr C(\lambda)\Bigr)
+\mathscr B(\lambda)\mathscr C(\mu),\label{br-12} \\
\nonumber \\
\{\mathscr B(\lambda),\mathscr C(\mu)\}_1&=&\frac{2}{\lambda-\mu}\Bigr(\mu \mathscr A(\mu)-\lambda \mathscr A(\lambda)\Bigl)+2\Bigl(1-\mathscr B(\lambda) \Bigr) \mathscr A(\mu),\nonumber \\
\nonumber \\
\{\mathscr C(\lambda),\mathscr C(\mu)\}_1&=&2\Bigl(\mathscr A(\mu)\mathscr C(\lambda)-\mathscr A(\lambda)\mathscr C(\mu)\Bigr).\nonumber
\end{eqnarray}
\end{prop}
\textbf{Proof}:
It is sufficient to check the statement on an open dense subset of the linear $r$-matrix algebra (\ref{rrpoi}) defined by the assumption that all the $h_i,e_i,f_i$ and $\epsilon_i$ are different.
Namely, substituting rational functions (\ref{abc1}) into (\ref{rrpoi})
one gets  canonical brackets on the direct sum of $n$ copies of  $sl(2)$
\begin{equation}\label{brhef-11}
 \{h_j,e_j\}_0=e_j,\qquad \{h_j,f_j\}_0=-f_j,\qquad\{e_j,f_j\}_0=2h_j,\qquad j=1,\ldots,n.
\end{equation}
Substituting these rational functions (\ref{abc1}) into the second brackets (\ref{br-12}) one gets  the following non-local brackets between generators $h_i,e_i,f_i$
\begin{eqnarray}
  \{h_j,e_j\}_1=(\epsilon_j-e_j)e_j,\quad \{h_j,f_j\}_1=-(\epsilon_j-e_j)f_j,\quad\{e_j,f_j\}_1=2(\epsilon_j-e_j)h_j, \nonumber\\
  \label{brhef-12}\\
  \{h_i,e_j\}_1=-e_ib_j,\quad \{h_i,f_j\}_1=e_ic_j,\quad \{e_i,f_j\}_1=-2e_ia_j,\quad
\{f_i,f_j\}_1=-2h_ic_j+2f_ia_j\,.\nonumber
\end{eqnarray}
Now it is easy to prove that Poisson bracket (\ref{brhef-11}) is compatible with the Poisson bracket (\ref{brhef-12}).

\begin{prop}
If
\begin{equation}\label{abc2}
\mathscr A=h_n\lambda+\sum_{i=1}^{n-1}  \frac{h_i}{\lambda-\epsilon_i},\quad
\mathscr B=e_n+\sum_{i=1}^{n-1}  \frac{e_i}{\lambda-\epsilon_i},\quad
\mathscr C=f_n+\sum_{i=1}^{n-1}  \frac{f_i}{\lambda-\epsilon_i}
\end{equation}
where $h_i,e_i,f_i$ are dynamical variables and $\epsilon_i$ are numerical parameters, then
the following brackets are compatible with linear $r$-matrix bracket (\ref{rrpoi})
\begin{eqnarray}
\{\mathscr B(\lambda),\mathscr B(\mu)\}_1&=&\{\mathscr A(\lambda),\mathscr A(\mu)\}_1=0,\nonumber \\
\nonumber \\
\{\mathscr A(\lambda),\mathscr B(\mu)\}_1&=&\frac{1}{\lambda-\mu}\Bigl(\lambda \mathscr B(\mu)-\mu \mathscr B(\lambda)\Bigr)-\rho_1 \mathscr B(\lambda)\mathscr B(\mu),\nonumber \\
\nonumber \\
\{\mathscr A(\lambda),\mathscr C(\mu)\}_1&=&\frac{-\lambda}{\lambda-\mu}\Bigl(\mathscr C(\mu)-\mathscr C(\lambda)\Bigr)
+ \rho_1 \mathscr B(\lambda)\mathscr C(\mu)-
  \rho_2 \mathscr B(\lambda),\label{br-22} \\
\nonumber \\
\{\mathscr B(\lambda),\mathscr C(\mu)\}&=&
\frac{2}{\lambda-\mu}\Bigl(\mu \mathscr A(\mu)-\lambda \mathscr A(\lambda)\Bigr)
+2\Bigl(1-\rho_1 \mathscr B(\lambda) \Bigr) \mathscr A(\mu)-\rho_3\,\mathscr B(\lambda)\nonumber \\
\nonumber \\
\{\mathscr C(\lambda),\mathscr C(\mu)\}&=&
-2\rho_1\Bigl(\mathscr A(\lambda)C(\mu)-\mathscr A(\mu)\mathscr C(\lambda)\Bigr)
+2\rho_2\Bigl(\mathscr A(\lambda)-\mathscr A(\mu)\Bigr)+\rho_3\,\Bigl(\mathscr C(\lambda)-\mathscr C(\mu)\Bigr)\nonumber
\end{eqnarray}
Here
\begin{equation}\label{rho1}
\rho_1=\frac{1}{e_n}=\left[ \frac{1}{\mathscr B(\lambda)}\right],\qquad \rho_2=\frac{f_n}{e_n}=\left[ \frac{C(\lambda)}{\mathscr B(\lambda)}\right],
\end{equation}
and
\begin{equation}\label{rho2}
\rho_3=1-\frac{2h_n(\lambda+\mu)}{e_n}+
\frac{2h_n\sum_{k=1}^{n-1} e_k}{e_n}=1- \left[ \frac{(\lambda+\mu)\mathscr A(\lambda)}{\lambda \mathscr B(\lambda)}\right]
-\left[ \frac{(\lambda+\mu)\mathscr A(\mu)}{\mu \mathscr B(\mu)}\right]\,,
\end{equation}
 where $[\mathscr X/\mathscr Y]$ is a quotient of polynomials $X$ and $Y$ in  variables $\lambda$ and $\mu$ over a field as in \cite{ts07a}.
\end{prop}
\textbf{Proof}:
As above it is sufficient to check the statement on an open dense subset of the linear $r$-matrix algebra (\ref{rrpoi}) defined by the assumption that all the $h_i,e_i,f_i$ and $\epsilon_i$ are different.

Namely, substituting rational functions (\ref{abc1}) into (\ref{rrpoi})
one gets local brackets $n-1$ copies of  $sl(2)$
\begin{equation}\label{brhef-21}
\{h_j,e_j\}_0=e_j,\qquad \{h_j,f_j\}_0=-f_j,\qquad\{e_j,f_j\}_0=2h_j,\qquad j=1,\ldots,n-1,
\end{equation}
and  degenerate brackets
\begin{equation}\label{brhef-21a}
\{e_n,f_n\}=-2h_n,\qquad \{h_n,h_i\}=\{h_n,e_i\}=\{h_n,f_i\}=0\,.
\end{equation}
The leading coefficients $h_n$ is the Casimir element for these brackets.

Substituting  rational functions (\ref{abc1}) into the second brackets (\ref{br-22}) one gets  second non-local brackets between generators $h_i,e_i,f_i$.
At $i,j=1,\ldots,n-1$ these brackets looks like
\begin{eqnarray}
\{h_j,e_j\}_1=\left(\epsilon_j-\frac{e_j}{e_n}\right)e_j,\quad \{h_j,f_j\}_1=-\left(\epsilon_j-\frac{e_j}{e_n}\right)f_j,\quad
\{e_j,f_j\}_1=2\left(\epsilon_j-\frac{e_j}{e_n}\right)h_j,\nonumber\\
\label{brhef-22}\\
\{h_i,e_j\}_1=-\frac{e_ib_j}{e_n},\quad \{h_i,f_j\}_1=\frac{e_ic_j}{e_n},\quad \{e_i,f_j\}_1=-\frac{2e_ia_j}{e_n},\quad
\{f_i,f_j\}_1=\frac{-2h_ic_j+2f_ia_j}{e_n}\,.\nonumber
\end{eqnarray}
At $e_n=1$ these brackets coincide with the previous brackets (\ref{brhef-12}).

The remaining non-zero brackets have the following form
\begin{equation}\label{brhef-22a}
\{f_i,f_n\}_1=\rho_3({\lambda+\mu=\epsilon_j})\,f_i,\qquad
\{e_i,f_n\}_1=-\rho_3({\lambda+\mu=\epsilon_j})\,e_i,\qquad
\{e_n,f_n\}=-e_n\,.
\end{equation}
Now it is easy to prove that Poisson bracket (\ref{brhef-21})-(\ref{brhef-21a}) is compatible with the Poisson bracket (\ref{brhef-22})-(\ref{brhef-22a}).
This completes the proof.

In the both cases we can rewrite second Poisson brackets in the following $r$-matrix form
\begin{equation}
\{\,\on{\mathscr L}{1}(\lambda),\,\on{\mathscr L}{2}(\mu)\}_1= [r_{12}(\lambda,\mu),\,
\on{T}{1}(\lambda)]- [r_{21}(\lambda,\mu),\on{T}{2}(\mu)\,]\,, \label{rrpoi2}
\end{equation}
where
\begin{equation}\label{r2}
r_{12}(\lambda,\mu)=\left(
                      \begin{array}{cccc}
                        \frac{\mu}{\lambda-\mu} & 0 & 0 & 0 \\
                        0 & 0 & \frac{\mu}{\lambda-\mu} & 0 \\
                        0 & \frac{\lambda}{\lambda-\mu} & 0 & 0 \\
                        0 & 0 & 0 & \frac{\mu}{\lambda-\mu}
                      \end{array}
                    \right)+\left(
                      \begin{array}{cccc}
                        0 & 0 & 0 & 0 \\
                        0 & 0 & \rho_1B(\mu) & 0 \\
                        -\rho_3 & 0 & 0 & 0 \\
                        \rho_2-\rho_1C(\mu) & 0 & 0 & 0
                      \end{array}
                    \right)
\end{equation}
and
\[
r_{21}(\lambda,\mu)= \Pi r_{12}(\mu,\lambda) \Pi.
\]
For the Lax matrix $\mathscr L$ with entries (\ref{abc1})  we have
\[
\rho_1=1, \qquad \rho_2=0,\qquad \rho_3=0.
\]
For the Lax matrix $\mathscr L$ with entries (\ref{abc2}) functions $\rho_k$ are given by (\ref{rho1})- (\ref{rho2}).

It is easy to see that entries of the Lax matrix (\ref{lax-a}) have the form (\ref{abc2}). So, we can use quadratic-linear algebra (\ref{rrpoi2}) in order to get bi-hamiltonian description
of the restricted multiple three wave interaction system. The first part of the brackets between variables $u,u^*$, $b_i,b_i^*$ and $c_i,c_i^*$ may be directly restored from the brackets (\ref{brhef-22})-(\ref{brhef-22a}). The remaining part has to be obtained from the compatibility conditions (\ref{sch-eq}).

As a result the non-zero brackets with variables $u$ and $u^*$ look like
\begin{eqnarray}
\{u,u^*\}_1=iu,\qquad
\{u^*,b_j\}_1&=-\frac{i}{2}b_j\rho_3(\lambda+\mu=\epsilon_j),\qquad
\{u^*,b^*_j\}_1&=\frac{i}{2}b^*_j\rho_3(\lambda+\mu=\epsilon_j),\nonumber\\
\{u^*,c_j\}_1&=\frac{i}{2}c_j\rho_3(\lambda+\mu=\epsilon_j),\qquad
\{u^*,c^*_j\}_1&=-\frac{i}{2}c^*_j\rho_3(\lambda+\mu=\epsilon_j).\nonumber
\end{eqnarray}
The local brackets at $j=1,\ldots,n$ are equal to
\[
\{b_j,c_j\}_1=\frac{ib_j^2}{2u},\qquad, \{b_j,c_j^*\}_1=\frac{i\epsilon_j b_j}{c_j},\qquad
\{b_j,b_j^*\}_1=\frac{i(b_jc^*_j-2u\epsilon_j)}{2u},
\]
\[
\{b^*_j,c_j\}_1=\frac{-i(b_jb^*_j+c_jc^*_j)}{2u},\qquad
\{b^*_j,c^*_j\}_1=\frac{-i(2b^*_j\epsilon_ju-c_j{c^*}^2_j)}{2c_ju},\qquad
\{c_j,c^*_j\}_1=\frac{-ib_jc^*_j}{2u}.
\]
The non-local brackets at $i\neq j$ read  as
\[
\{b_i,b^*_j\}_1=-\frac{ib_ic^*_j}{2u},\qquad \{b_i,c_j\}_1=\frac{ib_ib_j}{2u},\qquad
\{c_i,c^*_j\}_1=-\frac{ib_ic^*_j}{2u},\qquad \{c^*_i,b^*_j\}_1=-\frac{ic^*_ic^*_j}{2u},
\]
\[
\{c_i,c_j\}_1=\frac{i(b_ic_j-b_jc_i)}{2u},\qquad\{c_i,b^*_j\}_1=\frac{i(c_ic^*_j+b_ib^*_j)}{2u},\qquad
\{b^*_i,b^*_j\}_1=\frac{-(b^*_ic^*_j-c^*_ib^*_j)}{2u}.
\]
Other brackets are equal to zero, for instance
\[\{b_i,b_j\}_1=\{b_i,c^*_j\}=\{c^*_i,c^*_j\}_1=0.\]
Using $r$-matrix algebras (\ref{rrpoi}) and (\ref{rrpoi2}) it is easy to prove that integrals of motion $I_j\in\{J_0,J_i,\ldots,J_n,K_1,\ldots,K_n\}$  (\ref{int-mot}) are in the bi-involution with respect to the brackets $\{.,.\}_0$ and $\{.,.\}_1$:
\[
\{I_j,I_k\}_0=\{I_j,I_k\}_1=0\,.
\]
The one of the main problems is that the main characteristic of the model,  such as equation of motion,  the Lax matrices and integrals of motion are invariant with respect to conjugation
\begin{equation}\label{conj}
u\leftrightarrow u^*,\qquad b_j\leftrightarrow b^*_j,\qquad \qquad c_j\leftrightarrow c^*_j,
\end{equation}
but the second brackets $\{.,.\}_1$ do not invariant with respect to this transformation.

The second problem is that the second brackets $\{.,.\}_1$ are rational brackets and we have an obvious problem with quantization of these brackets.

According to \cite{ts07b} there are different bi-hamiltonian structures for a given integrable system. So, we could try to find another bi-hamiltonian description of the restricted multiple three wave interaction system, which allows as to avoid these problems.

\end{document}